\RequirePackage{fix-cm}
\documentclass[twocolumn,epjc3]{svjour3}  

\smartqed  % flush right qed marks, e.g. at end of proof
\RequirePackage{graphicx}
\RequirePackage{units}
\RequirePackage{cite}
\RequirePackage{amsmath}

\RequirePackage[colorlinks,citecolor=blue,urlcolor=blue,linkcolor=blue]{hyperref}
\RequirePackage{color}
\RequirePackage{multirow}
\providecommand{\tabularnewline}{\\}

\newcommand{\s}{E_{cm}}
\newcommand{\tev}{\textrm{TeV}}

\newcommand{\mgfive}{\textrm{MadGraph5\_aMC$@$NLO}}

\journalname{Eur. Phys. J. C}

\begin{document}

\title{
Going all the way in the search for WIMP dark matter at the muon collider through precision measurements
}

\titlerunning{WIMP searches at the high energy muon collider}        % if too long for running head

\author{Roberto Franceschini\thanksref{e1,addr1} and  Xiaoran Zhao\thanksref{e2,addr1} %etc.
}

\thankstext{e1}{e-mail:\href{mailto:roberto.franceschini@uniroma3.it}{roberto.franceschini@uniroma3.it}}
\thankstext{e2}{e-mail:\href{mailto:xiaoran.zhao@uniroma3.it}{xiaoran.zhao@uniroma3.it}}


\institute{Universit\`a degli Studi and INFN Roma Tre, Via della Vasca
Navale 84, I-00146, Roma \label{addr1}}

\date{Received: date / Accepted: date}

\maketitle

\begin{abstract}
Dark Matter is a necessary ingredient for a complete theory of Nature, which has so far  remained elusive in laboratory searches for new particles. Searches at current and future colliders are in principle a promising way to search for electroweak charged dark matter particles, but the  sensitivity of experiments at the LHC  and future $pp$ colliders falls short to fully probe the whole  mass range allowed for electroweak charged dark matter particles, which extends in principle up to the O(0.1) PeV.
In this work  we examine the effect of  on-shell and off-shell propagation of electroweak charged thermal dark matter particles on integrated and differential rates of several Standard Model final states at the muon collider, considering candidates 
from weak 2-plet at the TeV scale up to 7-plet and 9-plet in the O(0.1)
PeV ballpark. 
For fermionic WIMPs we find that all dark matter candidates with $n\leq 5$, corresponding to a thermal mass up to 14~TeV, can be excluded at the high-energy muon collider for some center-of-mass energy at or below 14~TeV. For the $n>5$ WIMPs our results show that  higher energy muon colliders offer a route to conclusively probe both scalar and fermionic WIMPs off-shell production all the way up to the perturbativity bound for WIMP dark matter at O(0.1)~PeV.
Our results  bring  WIMPs over the whole allowed mass range in the realm of collider searches and motivate research and development for the realization of a high energy muon collider.

\end{abstract}
\section{Introduction}

Some kind of new matter is required by several probes of physics at
the macroscopic scale in the present day Universe (e.g. \cite{Begeman:1989kf,doi:10.1146/annurev.astro.39.1.137})
as well as by probes of the Early Universe. What interactions are
felt by this new form of matter is a subject of investigation, as,
so far, observations require only gravity to be felt by this kind
of matter. Strikingly, the new form of matter needed to explain the
afore mentioned observations is required to not have any significant
electromagnetic interactions, so that it does not easily emit light
(if at all) and remains essentially inert in long-distance physics,
except for gravity. The lack of electromagnetic interactions for this
type of new matter has earned it the nickname ``dark matter''. 

Given this observational situation, dark matter currently is an ``umbrella''
term that embraces a large number of possible microscopic realization
of new matter, as well possible macroscopic objects, which can fit with gravitational observations. The large number of possibilities
that exist for possible realizations of dark matter has initiated
a number of experimental initiatives targeted to broad categories
of candidates (axions \cite{1510.07633v2}, WIMPS\cite{Goodman:1984dc},
extended objects such as roughly solar mass BH \cite{1603.00464v2,2008.10743v1,CHAPLINE1975_PBH},
galaxy-sized fields \cite{Hu:2000uq}...) as well as more specific
searches which may be sensitive only to a narrow category of candidates
(BH observables, e.g. superradiance \cite{Cardoso:2018aa,Brito:2015oca,Cardoso:2017sh,2004.12326v1},
fuzzy-dynamics in galaxies, and more \cite{Graham:2015cj,Khmelnitsky:2014aa,1911.11755v2,Blas:2019ab}).
In this work we will attack the question on the possible discovery
of Weakly Interacting Massive Particles (WIMP) dark matter by producing on-shell or off-shell
it in high-energy collisions at particle accelerators.

WIMPs emerge as a most simple dark matter candidate, as they are massive
particle carrying electroweak charge under the Standard Model $SU(2)\times U(1)$
gauge dynamics and they can be produced in the Early Universe by the
so-called thermal freeze-out mechanism~\cite{Lee_1977}. Thermal freeze-out
is a very attractive mechanism as it provides a very simple link between
the present day abundance of DM and the moment in the evolution of
the Early Universe in which the dark matter particles have stopped
being in thermodynamical equilibrium with the rest of the SM particles.
This moment of ``decoupling'' is in turn set by a Boltzmann
equation  and two microscopic parameters of the dark
matter: the mass and the strength of its interaction with the SM matter.
As strong interactions would pose a number of problems for a dark
matter candidate~\footnote{See e.g. Ref.\cite{DeLuca:2018mzn} and references there in for ways
out along this route.}, thermal freeze-out is ruled by the dark matter charge under the
$SU(2)\times U(1)$ gauge dynamics of the SM, and possible interactions of the dark matter with the Higgs boson.
Choosing an electroweak
representation, freeze-out dynamics provides a sharp prediction for the dark
matter mass, hence making the scenario very simple and predictive.

While the attention has long time been on the simplest WIMP candidates,
e.g. weak 2-plets and 3-plets appearing in supersymmetric models, the
landscape of WIMP dark matter candidates is much wider. Indeed, the
possibility to consider $n>3$ electroweak representations for the dark matter
candidates lends itself to use the general WIMP to build a catalog~\cite{2107.09688v1,Bottaro:2022aa}
of concrete and predictive dark matter scenarios that embrace the
whole big range of mass scales identified by the thermal freeze-out
\cite{1805.10305v2,Griest:1989wd} from the the 10 GeV scale to the
fraction of PeV. 

With such large range for WIMP dark matter mass it is impossible to
have a single experiment that can cover the whole range. Especially
for experiments at colliders, a great challenges resides in the possibility
that dark matter mass is well beyond the TeV scale. As collider experiments
attempt to produce the WIMP in the laboratory, the center of mass
energy required to produce the new particle can be prohibitively large.
The possibility to envision high energy lepton collider, e.g. muon
colliders~\cite{Black:2022ab,2203.07224v1,2203.08033v1,2203.07964v1,2203.07261v1,mucol2021,MICE2020,Delahaye:2019aa,Palmer_2014},
up to center of mass energies in excess of 10 TeV offers new possibilities
in this direction. This possibility is particularly exciting in view of the limited mass reach that even a futuristic 100 TeV $pp$ collider will attain due to reduced center of mass energy at which the proton constituents collide~\cite{Low:2014sh,Cirelli:2014ai,Di-Luzio:2018aa}.

In this work we study the potential of high energy muon colliders
to observe signals from dark matter candidates in SM measurements
of scattering processes such as 
\[
\mu^+\mu^-\to f\bar{f}+X\,,
\]
which has been previously considered in \cite{Di-Luzio:2018aa}
as well as new channels 
\[
\mu^+\mu^-\to f^{'}\bar{f}+X\,,
\]
where $f$ and $f'$ charges differ by one unit, hence the hard scattering
involves charged currents, and diboson final states
\[
\mu^+\mu^-\to Zh/W^{+}W^{-}+X\,.
\]

In our work we discuss the advantages of using final states $f$ for
which reliable charge and particle identification is possible so that
one can maximally exploit differential measurements. We also discuss
the gain that can be attained from releasing these stringent charge
and particle identification requirements and exploiting fiducial cross-section
measurements that are available for a larger set of detector objects. 

\section{The WIMP catalog}

The relic abundance from thermal freeze-out needs to be computed from
a detailed study of the rates $DM\,DM\leftrightarrow SM\,SM$ and other processes
that can convert $SM$ states into $DM$ states. For situations close
to the thermal equilibrium the relevant Boltzmann equations can be
simplified and a rough estimate for the relic abundance $\Omega_{DM}$ can be obtained according to 

\[
\Omega_{DM}\sim1/\sigma\sim M^{2}/C_{n,\textrm{eff}}\,,
\]
where $C_{n,\textrm{eff}}\sim n^{3}$ accounts for the weak charges of
the $n$-plet. This scaling is obtained barring bound states and Sommerfeld
enhancement, which gets a progressively worse approximation as $n$
grows. Indeed bound state formation and Sommerfeld enhancement make
grow the rates that keep the WIMP in equilibrium with the SM and the
resulting dark matter mass for large $n$-plet is significantly increased
by these physical effects\cite{Mitridate:2017fk}. 

In Table~\ref{tab:WIMP-thermal-mass} we report the masses that we
use in our calculations for each dark matter candidate. These are
taken from the latest calculations \cite{2107.09688v1} of WIMP thermal
masses for pure $SU(2)$ $n$-plets with zero hypercharge (Majorana
fermions and real scalars) or suitable pairs of fermions forming a
pseudo-Dirac fermion or a complex scalar combination whose interactions
with the $Z$ boson are easily suppressed as to evade present bounds
from direct searches of dark matter in ultra-low background experiments~\cite{1708.07051v1,1707.08145v1}
sensitive to weakly interacting particles.

\begin{table}
\begin{centering}
\begin{tabular}{c|c|c|c}
\hline 
$T_{W}$ & $Y$ & $M_{F}$/TeV & $M_{S}$/TeV\tabularnewline
\hline 
\hline 
2 & $\nicefrac{1}{2}$ & $1.1\pm0.1$  & $0.58\pm0.01$ \tabularnewline
\hline 
\hline 
\multirow{2}{*}{3}
& 0 & $2.86\pm0.01$ & $2.53\pm0.01$ \tabularnewline
\cline{2-4} \cline{3-4} \cline{4-4 } 
 & 1 &  
 $2.85\pm 0.14$
 & $2.12\pm0.05$ \tabularnewline
\hline 
\hline 
4 & $\nicefrac{1}{2}$ & $4.79\pm0.09$ & $4.98\pm0.05$ \tabularnewline
\hline 
\hline 
\multirow{1}{*}{5} & 0 & $13.6\pm0.8$  & $15.4\pm0.7$ \tabularnewline
\hline 
\hline 
\multirow{1}{*}{7} & 0 & $48.8\pm3.3$  & $54.2\pm3.1$ \tabularnewline
\hline 
\hline 
\multirow{1}{*}{9} & 0 & $113\pm15$ & $117.8\pm15.4$ \tabularnewline
\hline 
\end{tabular}
\par\end{centering}
\caption{\label{tab:WIMP-thermal-mass}Values of the WIMP thermal masses corresponding to $SU(2)\times U(1)_{Y}$
charges  from Ref.~\cite{2107.09688v1,Bottaro:2022aa}.}
\end{table}

It must be noted that upon mixing of different $n$-plets the dark
matter mass that fits the observed relic abundance becomes a function
of the mixing. In general, mixing a larger $n$-plet with a smaller
one reduces the mass necessary to reproduce the thermal relic abundance
compared to the case of pure $n$-plet. The reduced thermal mass
follows from the smaller $C_{n,\textrm{eff}}$ that characterizes the
mixed state. Such reduction of the mass might have a substantial impact
for collider phenomenology as it might make the difference between
having a dark matter candidate outside or inside the kinematic reach
for direct production at a specific collider. On top of reducing the
thermal mass, mixing with smaller $n$-plets reduces the expected
signal rate by a factor that closely tracks $C_{n,\textrm{eff}}$, as we
will see in the following Sections \ref{sec:Neutral-Currents} and
\ref{sec:Charged-Currents}. As a consequence it is not possible to
establish in full generality if it is easier to search for pure or
mixed state WIMPs. Still, we can conclude that the pure $n$-plet
is the most demanding candidate in terms of necessary center of mass
energy for direct production as well as for indirect effects that
decouple as $1/M^{2}$ or faster. With this spirit in mind we consider
the pure $n$-plet a sensible benchmark of the reach of colliders.
A detailed study would be necessary for the many possible mixed cases
and it is left for future work. 

\section{Neutral Currents\label{sec:Neutral-Currents}}
\begin{figure}[htbp]
\begin{center}
    \includegraphics[width=0.4 \linewidth]{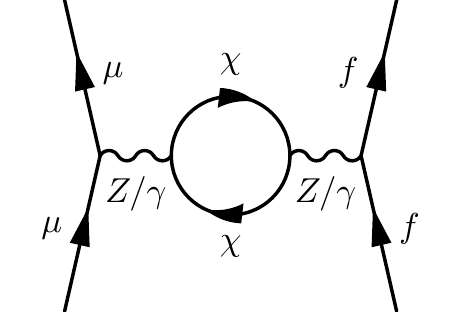}
    \includegraphics[width=0.4 \linewidth]{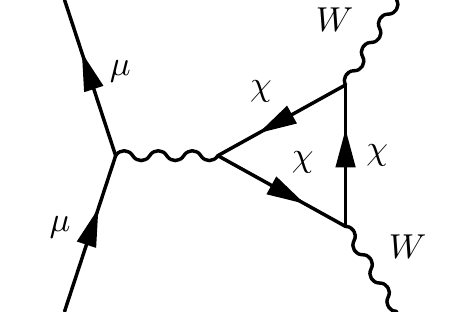}
\caption{{Neutral Current} diagrams from new electroweak matter $\chi$.}
\label{fig:nc}
\end{center}
\end{figure}

New electroweak matter affects the production rates of SM matter through new Feynman diagrams at 1-loop level, as shown in Fig.~\ref{fig:nc}. For fermion pair productions the new diagrams involve only bubble diagrams, whereas bosonic states, such as the $WW$ depicted in the figure, involve triangle loop functions. The different loop function structure and number of $\chi$ weak couplings make the two types of contribution rather distinct from each other. In both cases the effect of the new diagrams can be seen in modifications of both total   and differential rates for the production of the respective SM final state. Given the large amount of fermionic degrees of freedom in the SM the largest amount of observable information on the possible propagation of $\chi$ can be gained by understanding the behavior of the $f\bar{f}$ final state, on which we focus first.  

\begin{figure}[htbp]
\begin{center}
\includegraphics[width=0.99\columnwidth]{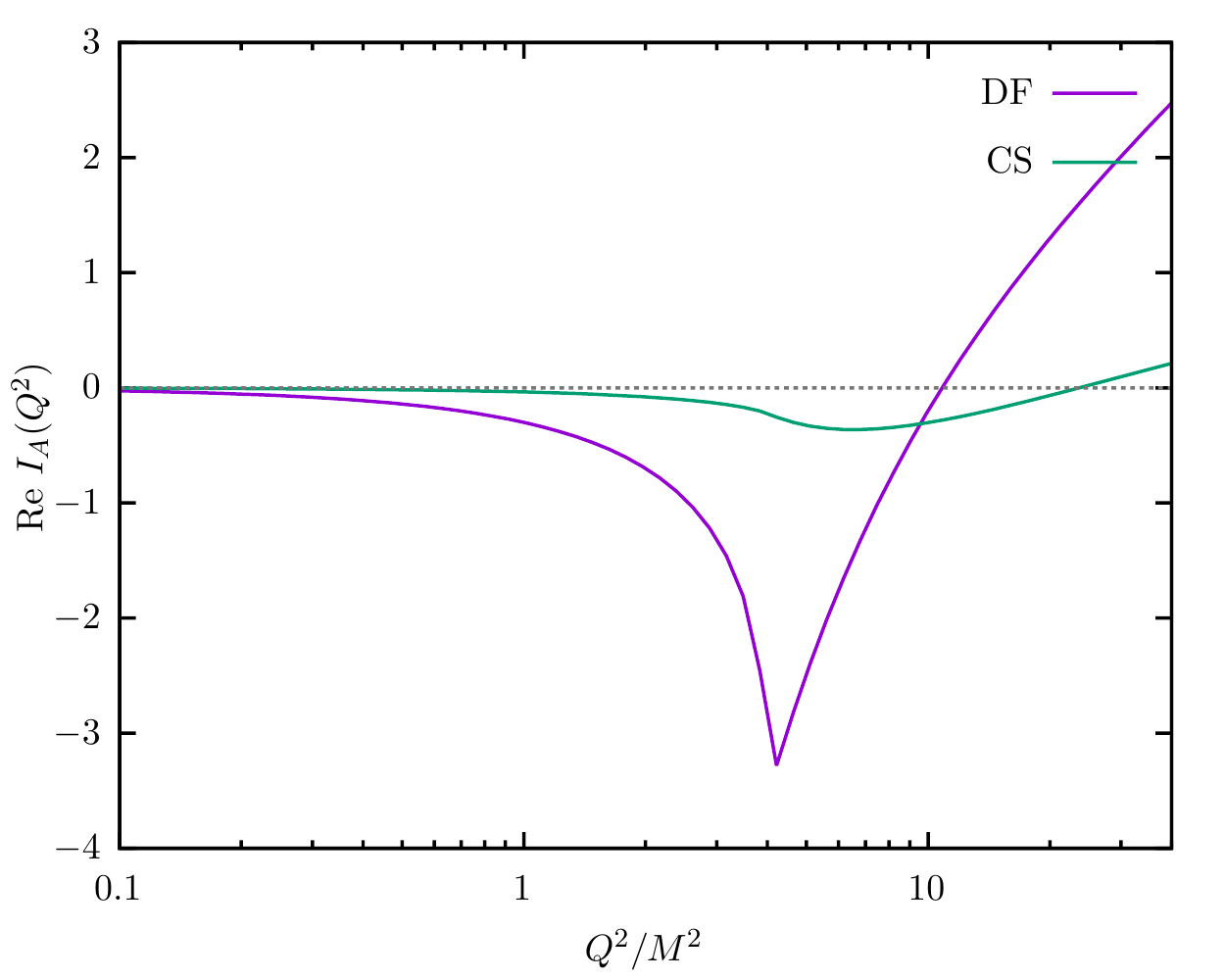}
\end{center}
\caption{The real part of the finite piece of the loop function for the self-energy correction with $\mu=M$.}\label{fig:loopfunc}
\end{figure}

For the  $f\bar{f}$ final state the leading effect from $\chi$ comes from interference between the first diagram in Fig.~\ref{fig:nc} and the SM amplitude for the same scattering. The contribution can be described by the corrections to the gauge boson self-energy, given by
\begin{align}
\Sigma^{VV}_T(Q^2)=-\frac{1}{16\pi^2}F_{VV}Q^2I_A(Q^2)\,,
\end{align}
where $I_A(Q^2)$ is the loop function that depends on the nature of the new matter,
and $F_{VV}$ is the collection of couplings between the vector boson and the new matter,
and their precise expressions are given in the appendix \ref{app:loopfunc}.
We note that $I_A(Q^2)$ is UV-divergent, and hence renormalisation is required to render it finite.
After renormalisation, only the finite piece of $I_A(Q^2)$ is relevant for our discussion. We plot the real part of the finite piece of $I_A(Q^2)$ in Fig.~\ref{fig:loopfunc}, as the imaginary part does not interfere with the SM tree-level amplitudes.
We can see that in the region $Q^2\ll M^2$, the contribution is suppressed and grows as $\mathcal{O}(Q^2/M^2)$, which can be well described in EFT.
As $Q^2$ increases, the Dirac Fermion contribution peaks at the threshold $Q^2=4M^2$, and decreases afterwards,
while the Complex Scalar contribution peaks at $Q^2\sim 6M^2$.
We note that for the Dirac Fermion with $Q^2\sim 10M^2$, and Complex Scalar with $Q^2\sim 24M^2$,
the real part is about zero, which implies that at this point the interference is strongly suppressed.
The channel $\mu^+\mu^-\to \mu^+\mu^-$ does not necessarily suffer this suppression as it also receives $t$-channel contribution at a different $Q^2$ region, and hence becomes important especially close to the zero of the $s$-channel. Concretely we will find that the sensitivity of new weak matter from the  $\mu^+\mu^-$ final states is  quite limited, so we will discuss possible remedies to this loss of sensitivity in Sec.~\ref{sec:Charged-Currents}.
We remark that in nearly all energy ranges the effects of Dirac Fermion are much larger than Complex Scalars,
which suggests that scalar dark matter is much more difficult to probe.

\begin{figure*}[htbp]
\begin{center}
\includegraphics[width=0.39\linewidth]{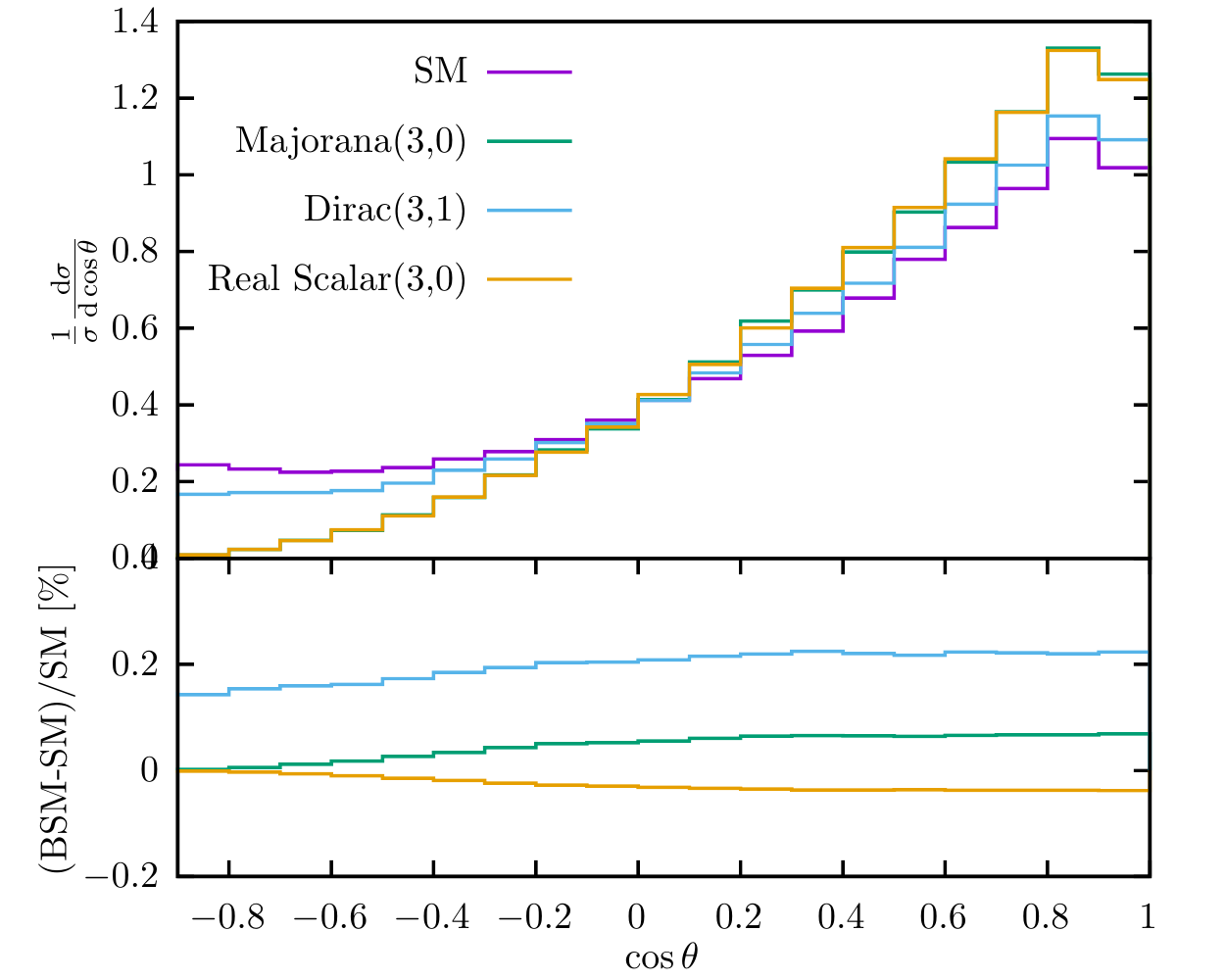}
\includegraphics[width=0.39\linewidth]{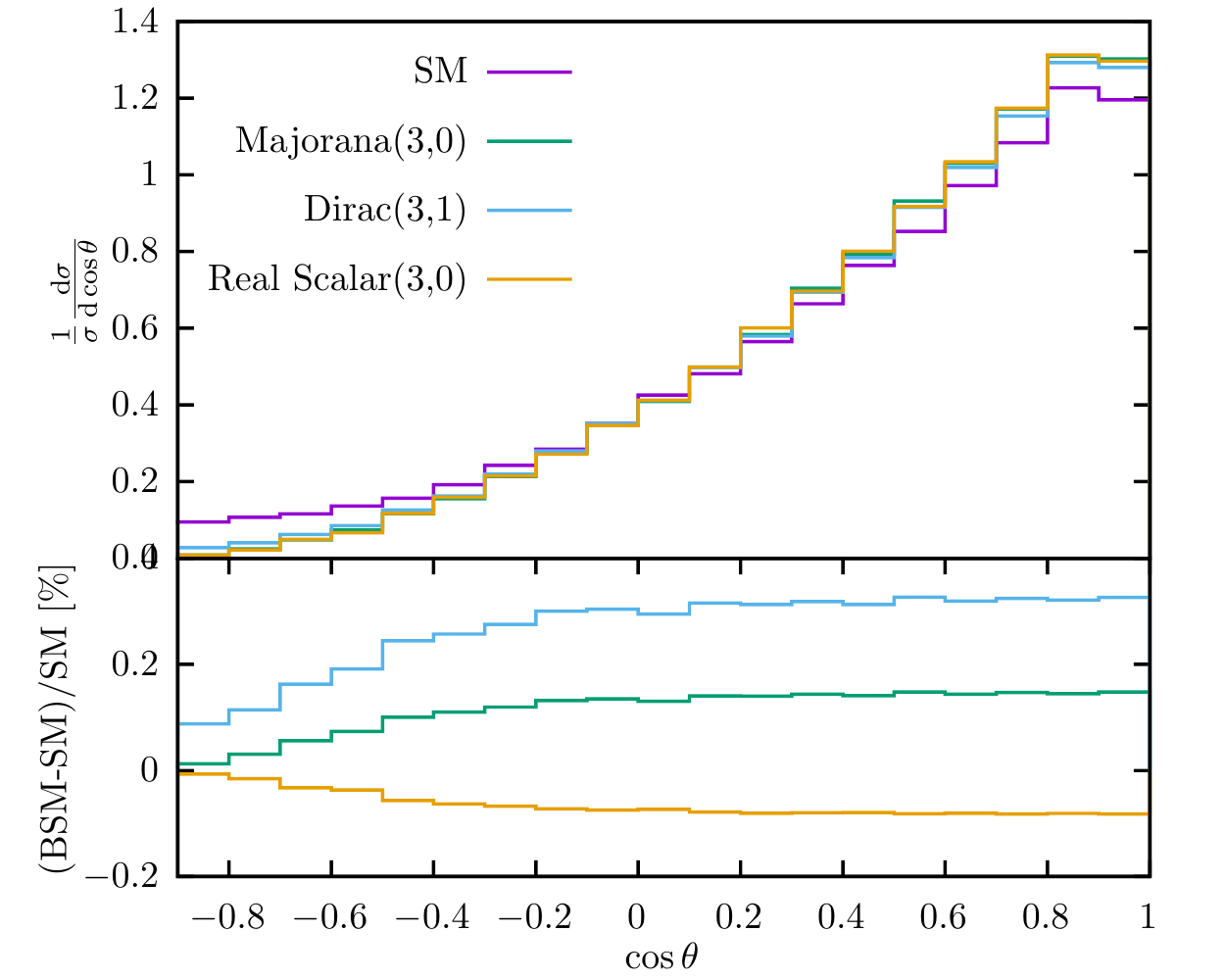}\\
\includegraphics[width=0.39\linewidth]{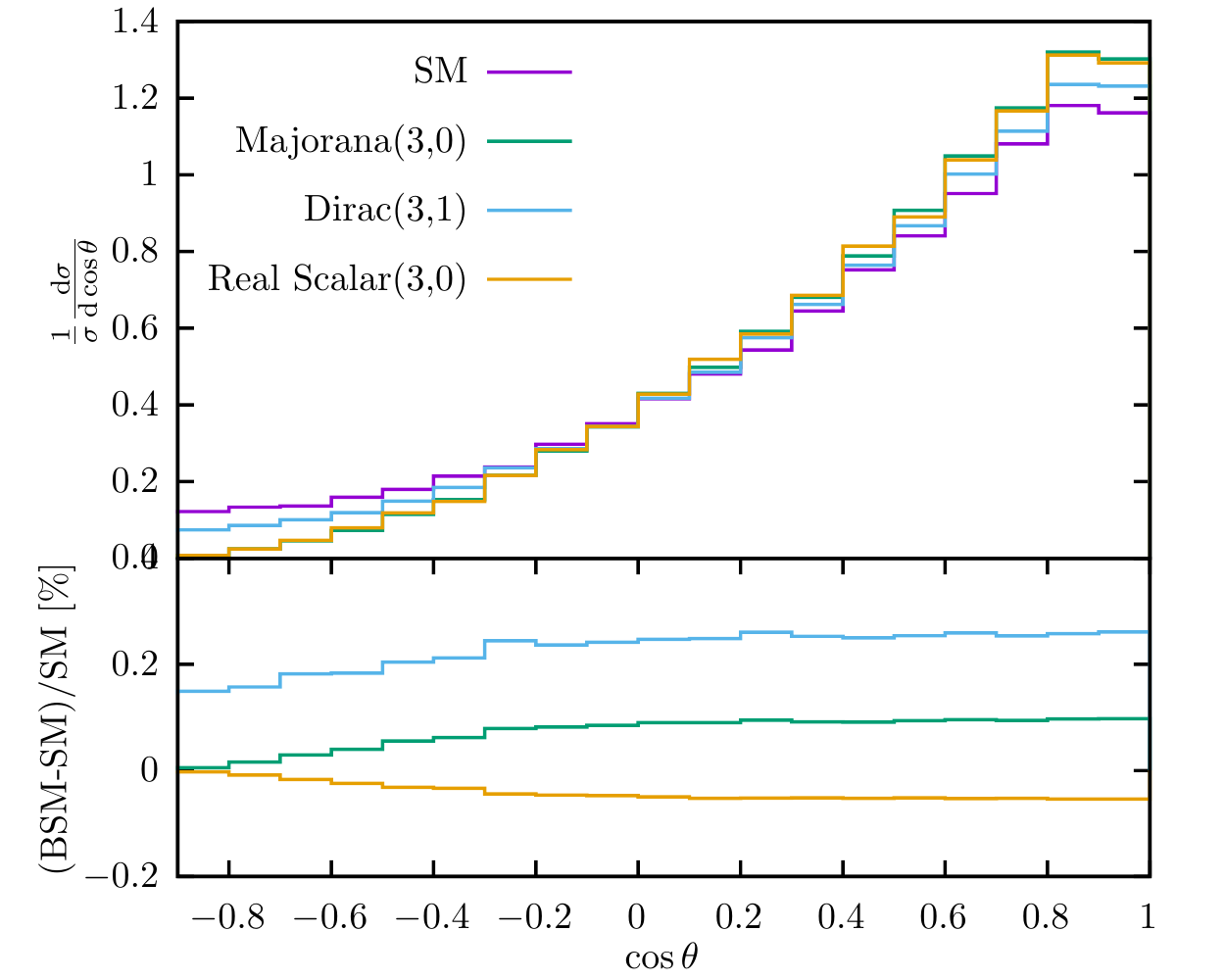}
\includegraphics[width=0.39\linewidth]{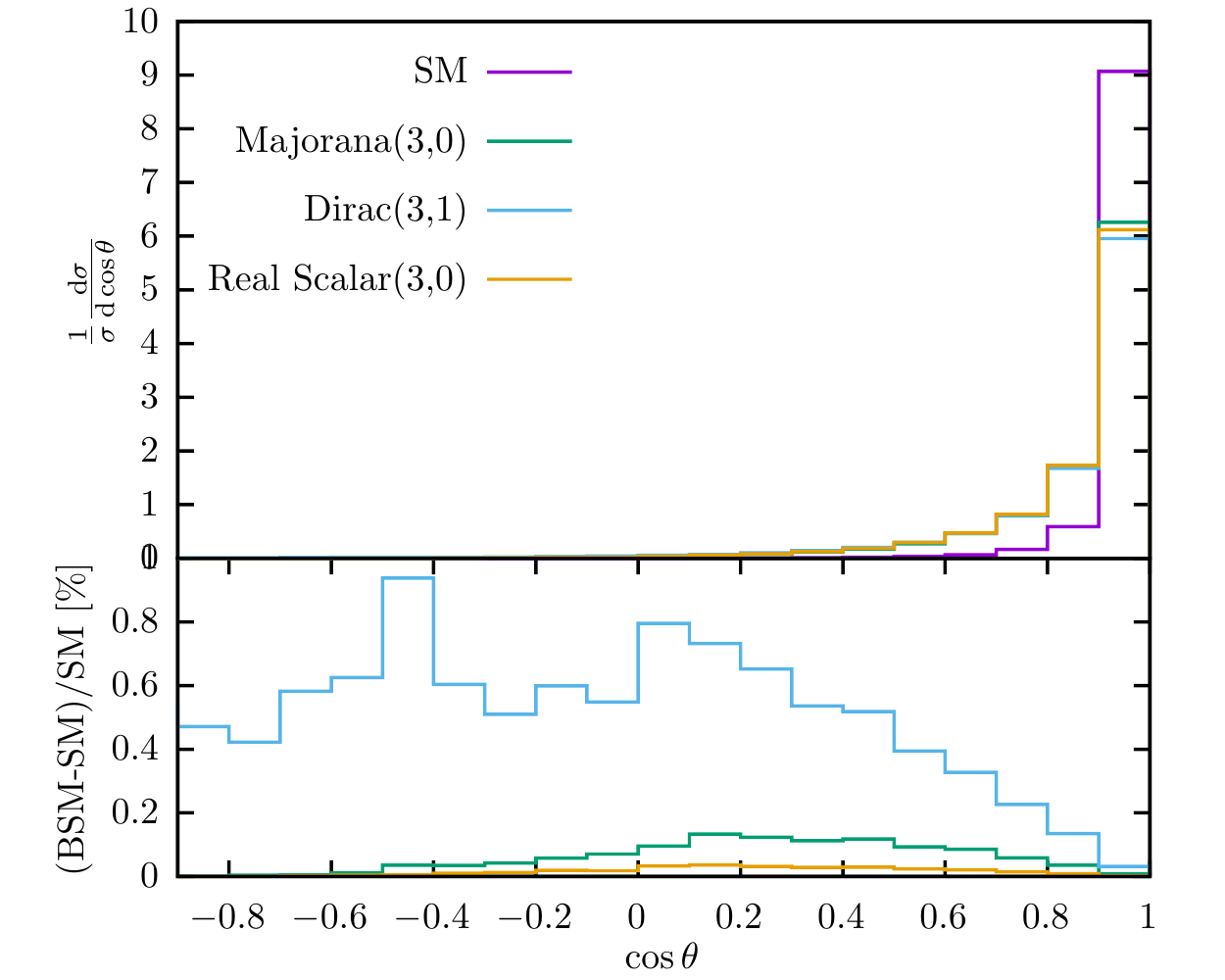}
\caption{{\bf Top panels} Polar angle distribution from the SM (purple cross), interference between diagrams mediated by complex scalar (cyan) or Dirac fermion (green) $\chi$ and the SM ones for the $e^{+}e^{-}$, $b\bar{b}$, $u\bar{u}$ and $\mu^{+}\mu^{-}$ final states at 10~TeV muon collider. {\bf Bottom panels} Ratio of the above for each fermionic final state.}
\label{fig:theta}
\end{center}
\end{figure*}

Another important feature of the new physics effect in neutral currents is that it does not depend on the initial and final state directly,
as it merely amounts to corrections to the gauge boson self-energy. If we consider the $g_1\to 0$ limit one can see that the SM and the BSM rates have the same dependence on the phase-space for a 2-to-2 scattering. To support the reasoning above we show in Fig.~\ref{fig:theta} the differential rate w.r.t. the polar scattering angle $\theta$ of the pure SM and of the interference of the BSM with SM amplitude for four different types of $f\bar{f}$, namely $f=u,b,e,\mu$~\footnote{$f=\nu$ is also possible, but in order to be observed it needs to be dressed with radiation of a $Z$ boson or a $\gamma$. This results in mono-$\gamma$ and mono-$Z$ signals, which have been studied at the muon collider in Refs.~\cite{2107.09688v1,2009.11287v1}. These final states have backgrounds dominated by SM processes involving missing momentum from particles escaping the detector acceptance. Such backgrounds are both very copious and not-interfering with our signal, thus leaving only the chance to observe the effects of $\chi$ not  via interference, but rather via the square of the BSM amplitudes, that are twice suppressed by the loop factors and the heavy mass of the $\chi$.  }, for both $Y=0$ and $Y\ne0$ dark matter candidates at {$\s=10$~TeV}. %
Except for the $\mu^{+}\mu^{-}$ channel, the cross-section of the SM and that of the leading BSM effect are distributed quite similarly over the range of polar angles, following the expectation for a scattering between left-handed currents, i.e. $\theta=\pi$ is distinctively suppressed. From this we observation we conclude that it is safe to ignore $Y$ to gain an understanding of the expected results. Still in our results we will always include the full effect from $Y$.

The above observation on the utility of the little distinctive power of the differential distributions has further consequences. Based on such observation, 
 we argue that channels containing charge-ID unfriendly jets, e.g. $d\bar{d},s\bar{s},u\bar{u}$ can be used rather effectively to probe the existence of $\chi$. These contributions have not been considered so far in the literature, which concentrated on the sensitivity from just charge-aware channels and their differential distributions (mainly $e,\mu$, and $b$). Given the large rate of jets these channels can significantly increase the statistics available for this study at the muon collider, thus speeding up discovery and making less of a bottleneck the luminosity that needs to be accumulated in order to gain sensitivity. The flip side of this strategy is that total rates are more sensitive than differential rates to dangerous systematic uncertainties, such as the luminosity measurement. Therefore in the discussion of our results we will try to delimit up to what point these total-rate observables can improve the sensitivity of the muon collider.

\section{Charged Currents\label{sec:Charged-Currents}}

At high energy lepton colliders with a center of mass energy $E_{cm}\gg m_V$,
the massive vector bosons are likely to be radiated from intial and final state particles.
In particular, such radiation is enhanced by the Sudakov logarithms $\frac{\alpha_W}{4\pi}\ln^2\frac{s}{m_V^2}$,
which can be of order 1 at large $E_{cm}$.
Thus, for a reliable theoretical prediction, such large logarithms should be resummed.

Fortunately, for the radiation of neutral particles, i.e. $Z$ and $\gamma$,
the KLN theorem guarantees that once we sum the real radiation and virtual corrections,
such large logarithms cancel.
As a result, the neutral current processes discussed in the previous section remain valid,
though they become inclusive processes rather than exclusive ones.
On the other hand, the radiation of the $W$ boson from initial state, leads to violation of KLN theorem.

The importance of charged current hard scattering is enhanced at larger
energy lepton colliders, as the possibility to radiate $W^{\pm}$
bosons is less and less suppressed as the energy transfer in the collisions
become much greater than the $W$ boson mass. These effects are of
primary importance in establishing high energy lepton collider reach
for several new physics scenarios \cite{2202.10509v1,2012.11555v1}
as they activate new channels. This type of effects highlights the
importance of electroweak corrections to analyses of high momentum
transfer scattering at high energy lepton colliders.

\begin{figure}[htbp]
\begin{center}
\includegraphics[width=0.30 \linewidth]{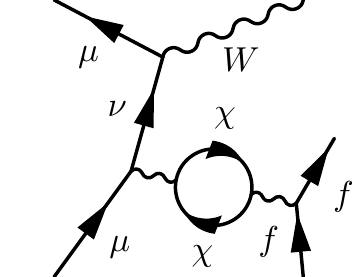}
\includegraphics[width=0.30 \linewidth]{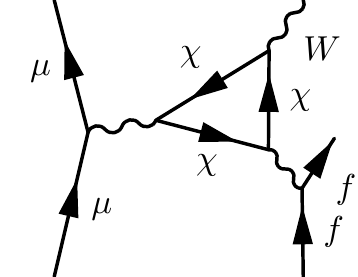}
\includegraphics[width=0.30 \linewidth]{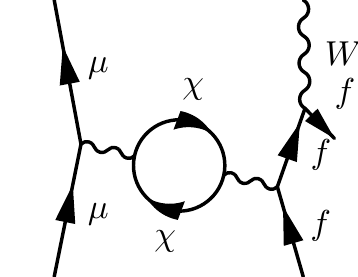}
\caption{Charged Current diagrams from new electroweak matter $\chi$}
\label{fig:cc}
\end{center}
\end{figure}

\begin{figure*}[htbp]
\includegraphics[width=0.49\textwidth]{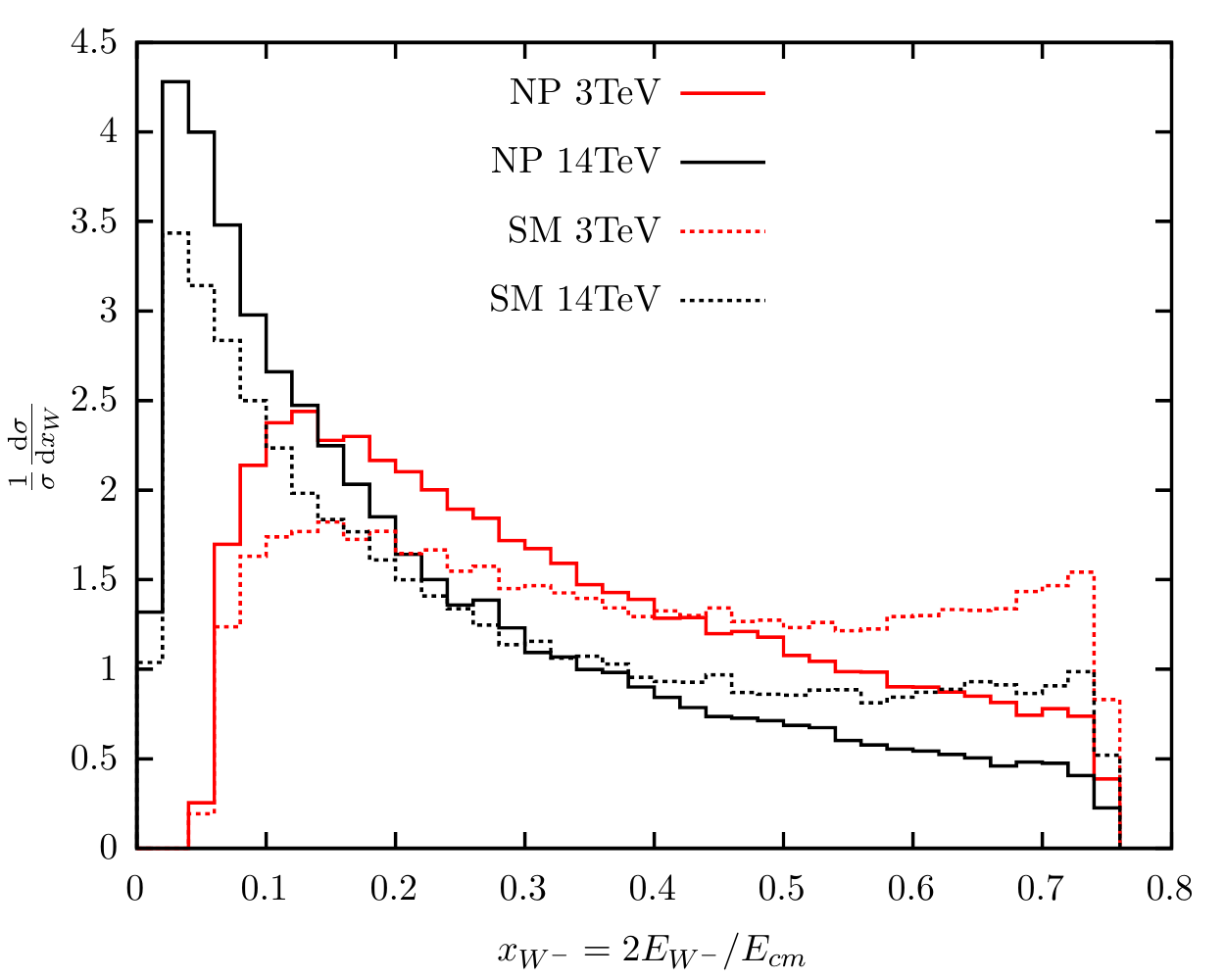}
\includegraphics[width=0.49\textwidth]{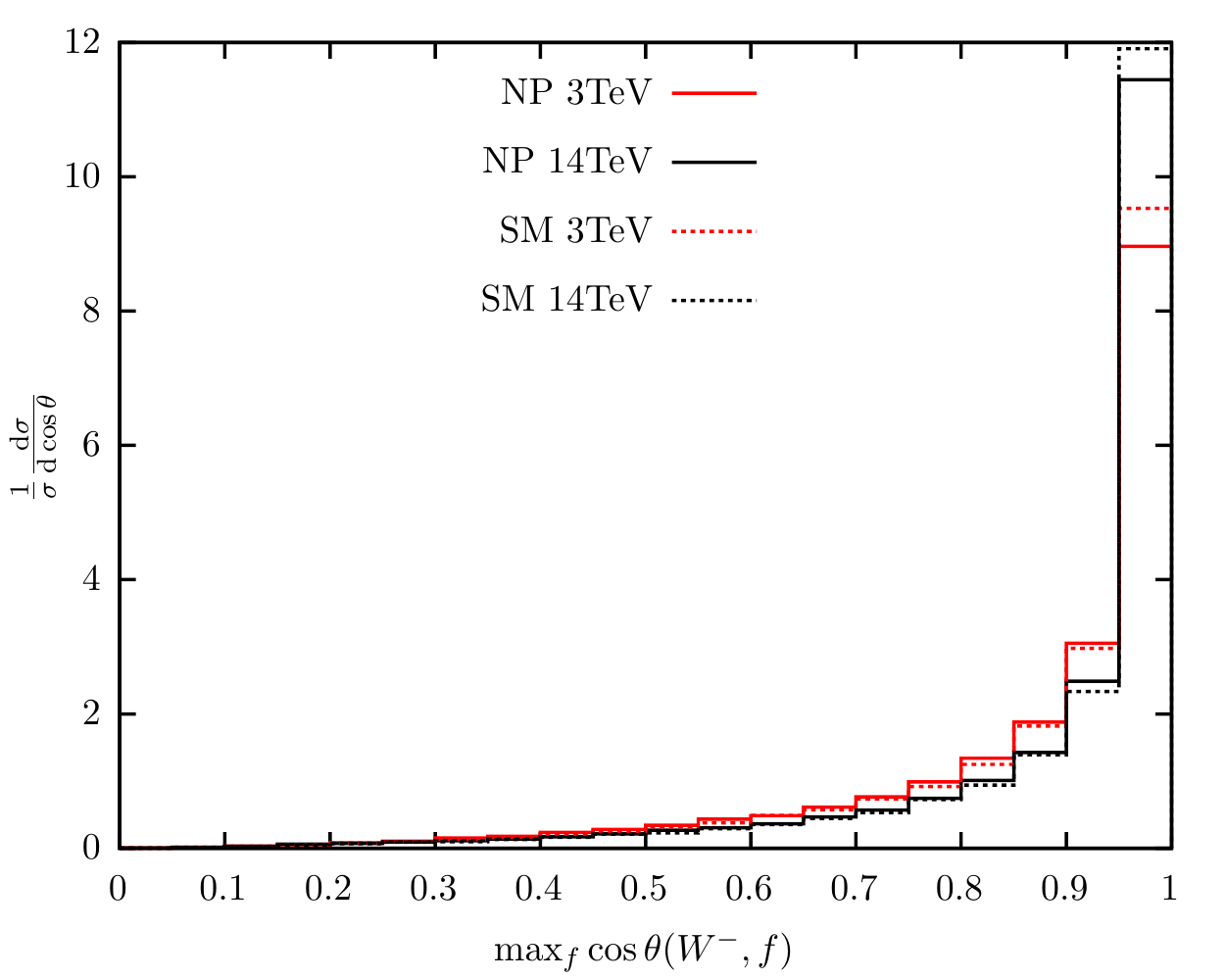}
\caption{Differential distribution of the normalized $W$ boson energy ($x_W=2E_{W}/E_{cm}$) and maximal $\cos\theta(W,f)$($f=\mu^+,\mu^-,u,\bar{d}$) for the $\mu^+\mu^-\to W^-u\bar{d}$ in the SM and the   interference  with a Majorana fermion 5-plet at the 3~TeV and 14~TeV muon collider.}\label{fig:difww}
\end{figure*}

In Fig.~\ref{fig:difww}, we show the differential distribution for the charged current process,
taking the $\mu^+\mu^-\to W^-u\bar{d}$ as an example for $\s=3$~TeV and {$\s=14$~TeV}. Results are shown for the SM and for the interference with a Dirac 5-plet, other WIMPs giving similar results.
In the left panel we show the distribution of the $W$ boson energy, normalized by the beam energy, $x_W=2E_{W}/E_{cm}$.
We can see that it is peaked around $x_{W}\to0$ for both signal and background,
and the $W$ boson becomes softer as the center of mass energy increases.
In the right panel, we show the maximal value of $\cos\theta(W,f)$, that is defined as the cosine of the minimal angle between $W$ and any fermion in the initial state and final state.
We can see that it is peaked around 1, which implies that the $W$ boson tends to be collinear to one fermion.

{We note that such soft/collinear $W$ boson radiation is expected for the high energy collisions that we study.} {This is similar to the radiation of photons and gluons that are enhanced in the soft/collinear regime, with the only difference that in our case  the $W$ boson  mass acts as the IR cut-off instead of a threshold for the detection of the photon or the jet activty stemming from the gluon.}

The impact of weak radiation in high energy muon collisions has been recently explored in Ref.~\cite{2202.10509v1}.
In particular it was shown that  large logarithms of the ratio of the hard scale of the process over cut-off scale around $m_{W}$ give $\mathcal{O}(1)$ corrections to the cross-sections of each process. Such corrections should be systematically resummed in order to retain good theoretical precision.
Furthermore, for a complete treatment of the radiation one would need to improve LO calculations by adding real radiation corrections for the neutral current channel and the necessary virtual corrections for each channel.
Such precise theoretical prediction is  not available at this time. Therefore we adopt a strategy to give predictions that are sufficiently accurate for our purposes using LO matrix elements, barring the loop of $\chi$ that we always consider explicitly.
To reach our goal we focus on the hard $W$ boson radiation, where the final state $W$ can be reliably separated from the rest of the final states and can be reliably treated by the LO matrix element of the  $\mu^+\mu^-\to W^{\pm}f\bar{f'}$ process.
In particular, we require that the $W$ boson carries a significant part of the total center-of-mass energy,
so that 
\begin{equation}
    0.5\cdot E_{cm}<m(f^{\prime}\bar{f})<0.9\cdot E_{cm}.\label{eq:CC1}
\end{equation}
Furthermore, we require that the $W$ boson is in the detector acceptance by  requiring 
\begin{equation}\label{eq:CC2}
8^\circ<\theta(W)<172^\circ\,,
\end{equation}
similarly to our selection for the NC channel explained in later Section~\ref{sec:mass-reach}. In our study we  only consider hadronic and semi-leptonic final state,
i.e. only $\mu^+\mu^-\to W^{\pm}(\to jj)e^{\mp}\nu$ and $\mu^+\mu^-\to W^{\pm}(\to e^{\pm}\nu,\mu^{\pm}\nu,jj)jj$ are included in our analysis.

\begin{figure}[htbp]
\begin{center}
\includegraphics[width=0.99\linewidth]{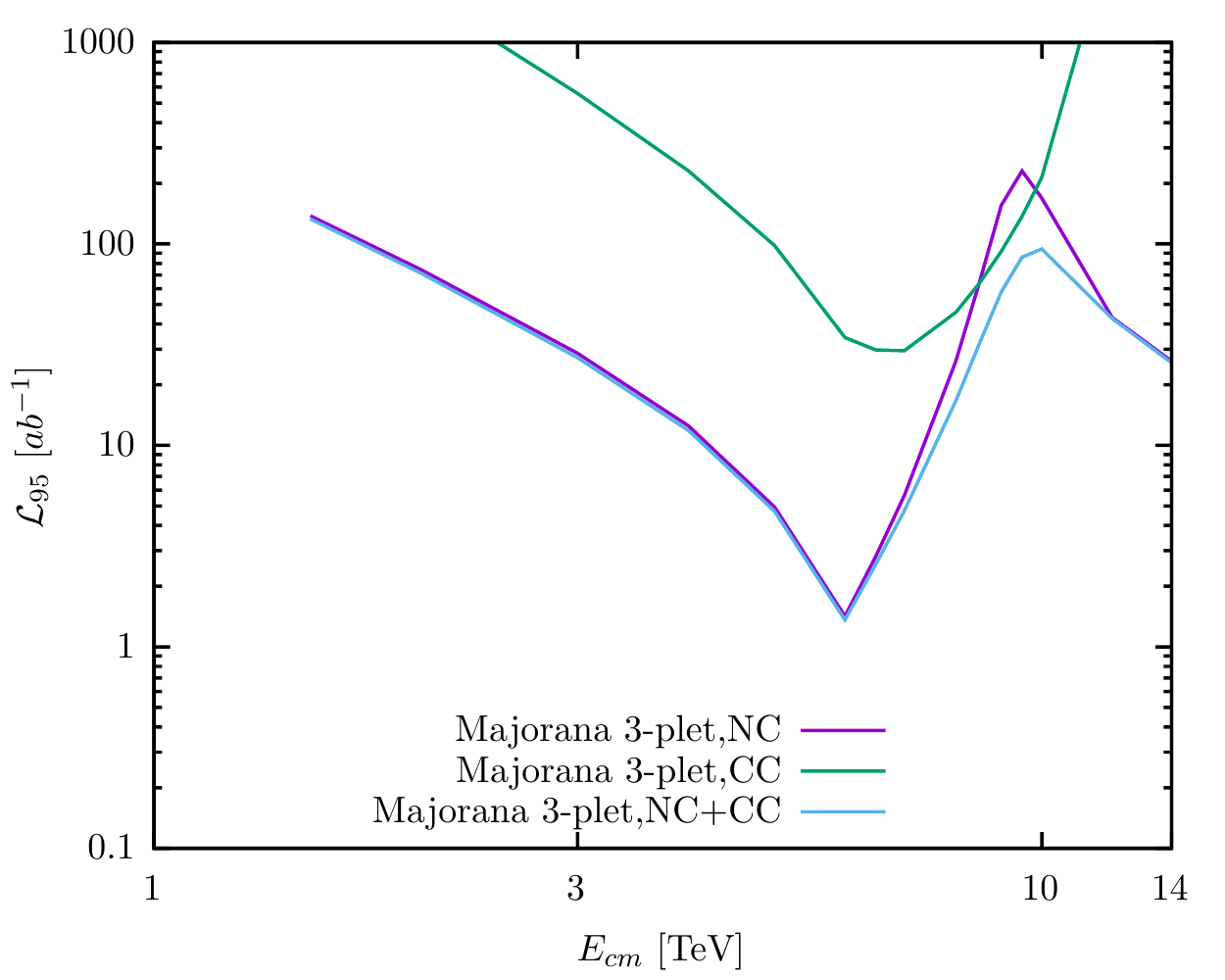}
\end{center}
\caption{Comparison of the Neutral Current(NC) channel and Charged Current(CC) channel for Majorana Fermion triplet.}\label{fig:cc-comp}
\end{figure}

In Fig. \ref{fig:cc-comp}, we show the importance of the CC channel, by comparing it with the NC channel, taking the Majorana Fermion with $n=3$ as an example.
The DM mass is 2.86 \tev, and we can see that for $E_{cm}<2M=5.76~\tev$, the required luminosity decreases as the energy increases,
and reaches the minimal value at the threshold $E_{cm}=2M$.
In such region, due to limited cross section, the importance of the CC channel defined as eqs.(\ref{eq:CC1})-(\ref{eq:CC2}) is negligible compared to the NC channel.
As $E_{cm}$ increase the real part of the interfering $s$-channel BSM amplitudes decreases and  reaches a zero around $E_{cm}\sim9~\tev$, thus leaving only the $t$-channel effects through $\mu^+\mu^-$ final state to provide a very loose constraint from NC processes.
On the other hand, for the CC channel the $Q^2$ is reduced due to the $W$ radiation and
the real part of the loop function $I_A(Q^2)$ is no longer zero. Under these conditions the CC provides better constraints than NC channel and contributes significantly to the overall sensitivity to $\chi$ from the precision measurements that we consider at the muon collider.

\section{Mass reach on thermal WIMPs \label{sec:mass-reach}}

To derive results from the general calculations introduced in the above Sections~\ref{sec:Neutral-Currents} and \ref{sec:Charged-Currents} we mimic experiments condition by considering  detector acceptance and efficiencies explained in the following. {All the calculations have been carried out using a customized model of \mgfive\,~\cite{Alwall:2014hca} that allows to deal with $\chi$ loops as described in \ref{app:loopfunc}.} 
The actual implementation of the effects of $\chi$ in a \mgfive\, model is based on the generation of one-loop triangle diagrams using Qgraf\cite{Cullen:2011ac} and process them through FORM\cite{Vermaseren:2000nd}.
The one-loop integrals are reduced to scalar integrals using Kira~\cite{Maierhofer:2017gsa}.
The one-loop scalar integrals are evaluated by LoopTools~\cite{Hahn:1998yk}. Besides the triangle diagrams, we also included bubble diagrams, as well as relevant counter terms introduced through renormalisation.

    For all channels, we consider as visible particles only those in the polar angle range 
    \begin{equation}
    8^\circ<\theta<172^\circ\,.\label{eq:acceptance}
    \end{equation}
We remark that for simplicity for  the channels $\mu^+\mu^-\to t\bar{t},W^+W^-,HZ$ we apply the angular requirement directly on $t$, $H$, $Z$ and $W$ particles instead of their decay products.

Unless noted otherwise we use fiducial cross-sections as the observables for our search of new physics effects. This is motivated by our findings in earlier sections and in particular by the  discussion around Fig.~\ref{fig:theta} on the shape of the BSM angular distributions.
For each channel we note the following specifications:
\begin{itemize}
    \item[$\circ$] for the $jj$ channel  we include both light flavors $j=u,d,s$ and heavy flavors $j=c,b$ without any tagging assuming 100\% efficiency;
    \item[$\circ$] for the $e^+e^-$ channel we assume 100\% efficiency;
    \item[$\circ$] for $\tau^+\tau^-$, we apply $\tau$-tagging with efficiency 50\%, and only consider events where both $\tau$ are tagged;
    \item[$\circ$] for $\mu^+\mu^-$, we study the differential distribution in the polar scattering angle using twenty bins, distributed equally on $-1<\cos\theta<1$ and we assume 100\% efficiency;
    \item[$\circ$] for $t\bar{t}$ final state we consider only hadronic decays ($t\bar{t}\to b\bar{b}jjjj$) and semi-leptonic decays ($t\bar{t}\to\ell\nu b\bar{b}jj$);
    \item[$\circ$] for $W^+W^-$ we consider only hadronic decays ($W^+W^-\to jjjj$) and semi-leptonic decays ($W^+W^-\to\ell\nu jj$);
    \item[$\circ$] for $HZ$ we consider only $H\to b\bar{b}$ and $Z\to jj,\ell^+\ell^-$ with $\ell=e,\mu\,$.
\end{itemize}

\begin{figure*}[htbp]
\begin{center}
\includegraphics[width=0.49 \linewidth]{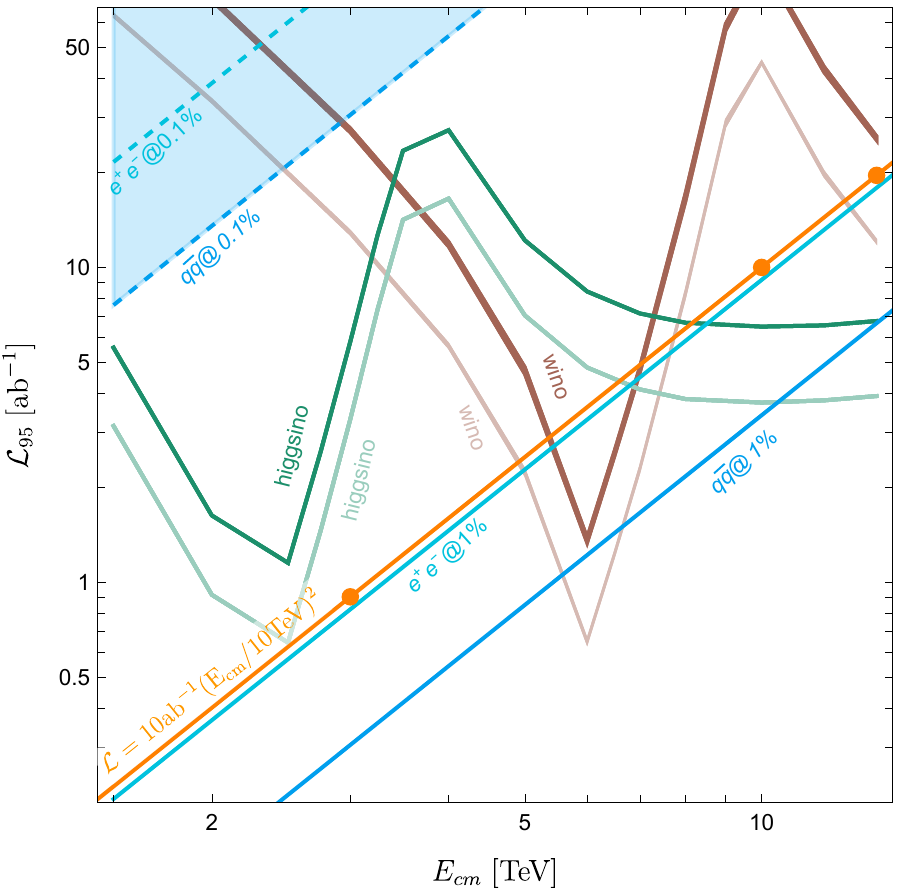}
\includegraphics[width=0.49 \linewidth]{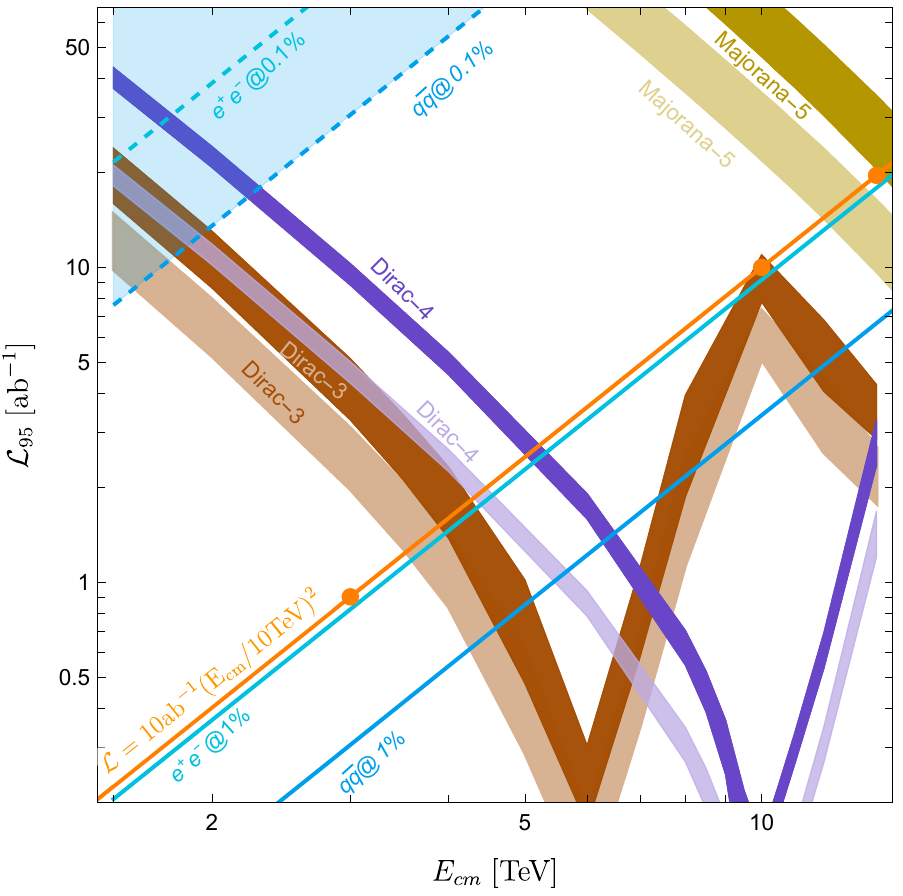}
\caption{Luminosity needed for a 95\% CL exclusion at a given center-of-mass energy $E_{cm}$. Each dark matter candidate is labelled next to its line or band. Bands thickness reflect the uncertainty on the thermal mass given in Tab.\ref{tab:WIMP-thermal-mass}. Lighter color lines and band correspond to polarized beams. The orange line tracks the expected baseline luminosity at each energy of the high energy muon collider. Any dark matter candidate whose line goes below the orange line can be excluded at the high energy muon collider for the corresponding values of $E_{cm}$.}
\label{fig:bounds}
\end{center}
\end{figure*}

Putting together the effect of $\chi$ on charged and neutral currents we obtain an inclusive result on the expected modifications of the fiducial rates and distributions for each given mass and weak charge. In Fig.~\ref{fig:bounds} we present the minimum required luminosity to put a 95\% CL bound on the existence of electroweak matter suitable to be dark matter at a muon collider running at center-of-mass energy $E_{cm}$. Each line in the figure has a corresponding darker and lighter color result, which give the results for unpolarized and polarized beams, respectively. 
The polarized beams are chosen as to maximize the left-handed component, that is more sensitive to  $\chi$. We pick 30\% left-handed polarization for both the positive and negative charge beams as to maximize the left-handed currents scattering. For those particles listed in Tab.~\ref{tab:WIMP-thermal-mass} with a sizable uncertainty on the thermal mass we draw bands, instead of lines, as to cover each respective thermal mass range. The results shown take into account the concrete value of $Y$ of each dark matter candidate. However the results are very similar to those that can be obtained by removing the $Y$ contribution in the BSM amplitudes.

In Fig.~\ref{fig:bounds} the orange line tracks the luminosity expected for a muon collider operating at center-of-mass energy $E_{cm}$, therefore all the dark matter candidates whose line drops below the orange line can be excluded at the high-energy muon collider in the corresponding range of values of $E_{cm}$. The orange line corresponds to about the luminosity necessary to measure the neutral current rate for $e^{+}e^{-}$ at 1\% precision. With such luminosity the dijet rate, owing to a much larger total rate, will be measured in deep sub-percent precision, but still probably not precisely enough to require a very careful analysis of systematic uncertainties. For reference in  Fig.~\ref{fig:bounds} we draw the line that corresponds to the luminosity necessary to measure the dijet rate at 0.1\% precision and we shade all the part of the plot above this line, as to indicate that luminosities above that line are so large that even tiny sources of uncertainty need to be evaluated before claiming sensitivity to dark matter. 

We observe that all the fermionic dark matter candidates with $n\leq 5$ can be excluded at the high-energy muon collider for some center-of-mass energy at or below 14~TeV using the baseline luminosity. The higgsino-like 2-plet, a notoriously elusive dark matter candidate, can be excluded at low energy, close to its production threshold around 2~TeV, only if beams can be polarized. Otherwise a collider at $E_{cm}\geq 8$~TeV, well above the threshold energy for 2-plet pair production is needed. The possibility to probe a Majorana 5-plet at 14~TeV also seems to hinge on the availability of polarized beams if one stick very strictly to the baseline luminosity. In absence of polarization, otherwise, the luminosity required for an exclusion may be slightly larger than the baseline. 

It should be remarked that for the $n=5$ Majorana fermion the pair production threshold for thermal mass is around 28~TeV, thus the effects of the $5$-plet WIMP can be captured approximatively in an EFT expansion over the parameter $Q^2/4M^{2}$, e.g. via the measurement of the $W$ parameter~\cite{Barbieri:2004ek}. Putting together the results of Ref.~\cite{Cirelli:2005uq,Bottaro:2022aa} on the size of the $W$ parameter generated by the Majorana 5-plet and those of Ref.~\cite{2202.10509v1} on the sensitivity of the muon collider to the $W$ parameter from measurements of NC and CC processes, we find that our result is in overall agreement with what can be obtained from these references. Our results is nevertheless slightly stronger than what can be cast from these references due to larger polar angle coverage being considered in our analysis. 

It is worth noting that the comparison of our result with that of Ref.~\cite{2202.10509v1} requires some care. In particular in our treatment the distinction between NC and CC at fixed leading order underestimates the importance of CC in setting bounds compared to a calculation that includes resummation of weak radiation as in Ref.~\cite{2202.10509v1}. We checked that the combined limit obtained from our procedure from NC and CC agrees well with the resummed result when same acceptance and same  event selection criteria are used. This detailed comparison and the possible  disagreement on the importance of each exclusive channels witnesses the need for further study of weak radiation at the high-energy muon collider.  

\begin{figure*}[htbp]
\begin{center}
\includegraphics[width=0.49 \linewidth]{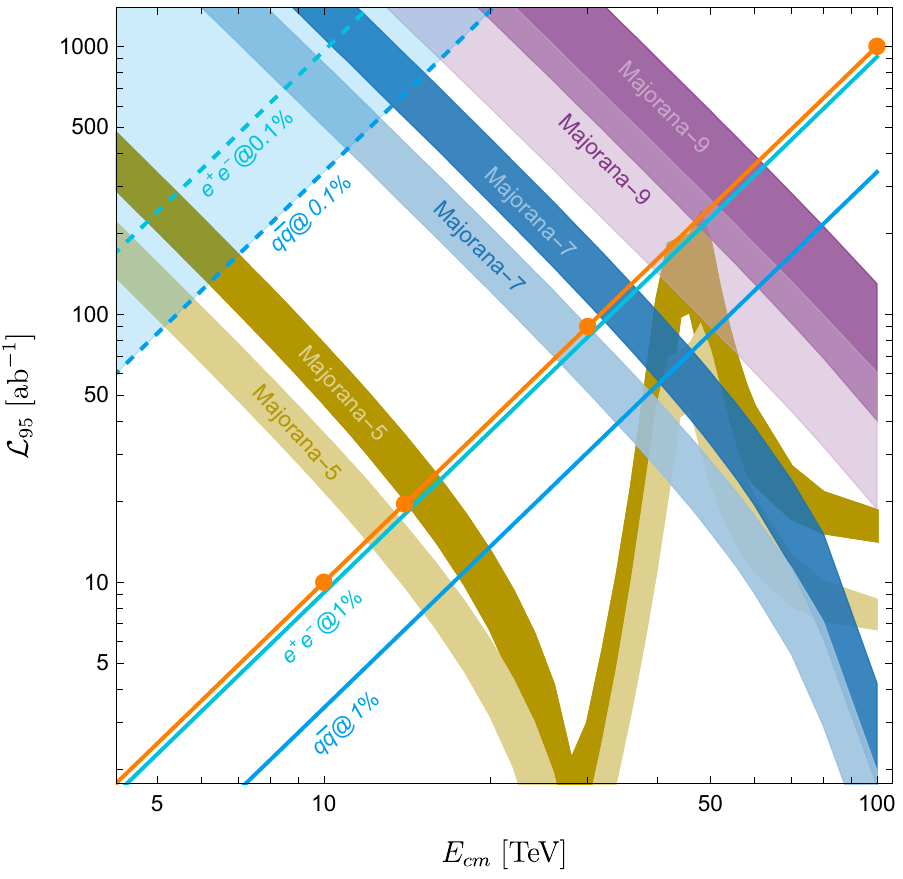}
\includegraphics[width=0.49 \linewidth]{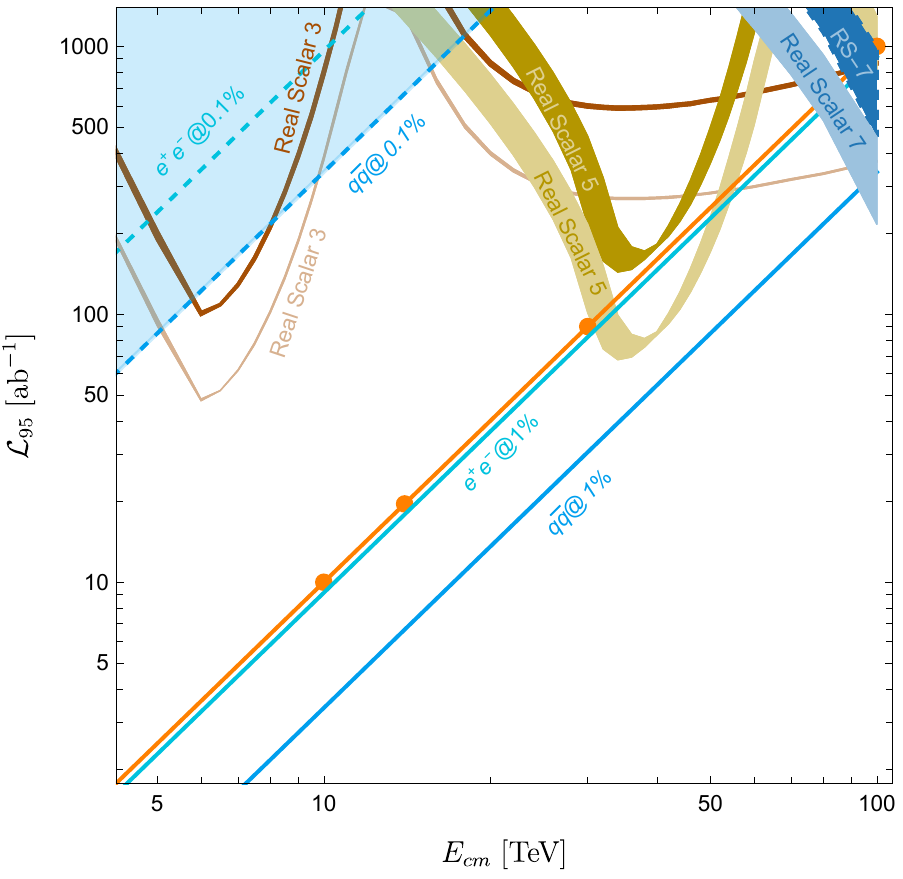}
\caption{Luminosity needed for a 95\% CL exclusion at a given center-of-mass energy $E_{cm}$. Each dark matter candidate is labelled next to its line or band. Bands thickness reflect the uncertainty on the thermal mass given in Tab.\ref{tab:WIMP-thermal-mass}. Lighter color lines and band correspond to polarized beams. The orange line tracks the expected baseline luminosity at each energy of the high energy muon collider. Any dark matter candidate whose line goes below the orange line can be excluded at the high energy muon collider for the corresponding values of $E_{cm}$.}
\label{fig:bounds-largen-scalars}
\end{center}
\end{figure*}

While the above result calls for further study to gain more control on the predictions of BSM effects once weak radiative corrections become relevant for 10+~TeV muon colliders, it is possible to look at higher energy colliders both with our explicit computation and using the scaling of EFT effects such as the $W$ parameter generated by higher $n$-plets. As a matter of fact we find that the reach of the muon collider can be extended further to probe more WIMP candidates as one considers larger $n$ and progressively larger $\s$. Figure~\ref{fig:bounds-largen-scalars} shows the required luminosity for a 95\% CL exclusion of 5, 7, and 9-plet Majorana dark matter candidates.  E.g. for the Majorana $n=7$ dark matter candidate, whose pair  production threshold is around 100~TeV, we find that a $\s=$30~TeV muon collider can measure the processes that we have considered and extract a $1\sigma$ measurement of the $W$ parameter at a precision around {$0.15 \cdot 10^{-7}$}, which probes at 95\% the effect expected around $0.3\cdot 10^{-7}$ from this dark matter candidate. Such dark matter candidate has a weak charge so large that its scattering rates at the LO in perturbation theory fill about few \% of the maximum rate allowed by perturbation theory (see \cite{2107.09688v1,Bottaro:2022aa}). That is to say that this dark matter candidate starts to exhibit a perturbation theory expansion that is all but merely ``perturbed'' by the next order in the expansion. 

At the present time the SM augmented by a Majorana 7-plet is a sensible and reasonably computable theory. However, it is fair to say that larger $n$-plets such as Majorana 9-plets and 11-plets have less interest in the context of WIMPs, as their charges are so large that perturbation theory converges very slowly and a Landau pole emerges within very close range to their mass~\cite{2107.09688v1}. With these provisions in mind, we can say that a muon collider program, if able to reach center-of-mass energy around 30~TeV, will be able to definitively probe fermionic WIMP candidates in the perturbative regime. In addition to this ``closure'', and very remarkably for the development of the machine, each stage of the collider at lower energy has a great potential to probe conclusively one or more dark matter candidates. 

For scalars we show results in the right panel of Fig.~\ref{fig:bounds-largen-scalars}. The reduced number of degrees of freedom in a scalar field makes the effects of scalar DM significantly smaller than for fermions. As a consequence it is necessary to run for luminosities much larger than the baseline in order to have a chance to put a constraint on scalar dark matter. The results in the figure correspond to real scalars, hence can be said to be somewhat on the pessimistic side, because this kind of field contain the minimal possible number of degrees of freedom and has the largest possible thermal mass for a given $SU(2)$ charge. Nevertheless we note that for large $n$ the effects due to the large charge overtakes the growth of the thermal mass and it is in priciple possible to test a real scalar 5-plet and  a 7-plet  at the baseline luminosity for center of mass energy 30 and 100~TeV, respectively. Though very large, these center-of-mass energies are not inconceivably large and it is nice to see that otherwise very elusive dark matter candidates can in principle be probed at colliders.

\begin{figure}[htbp]
\begin{center}
\includegraphics[width=0.99 \linewidth]{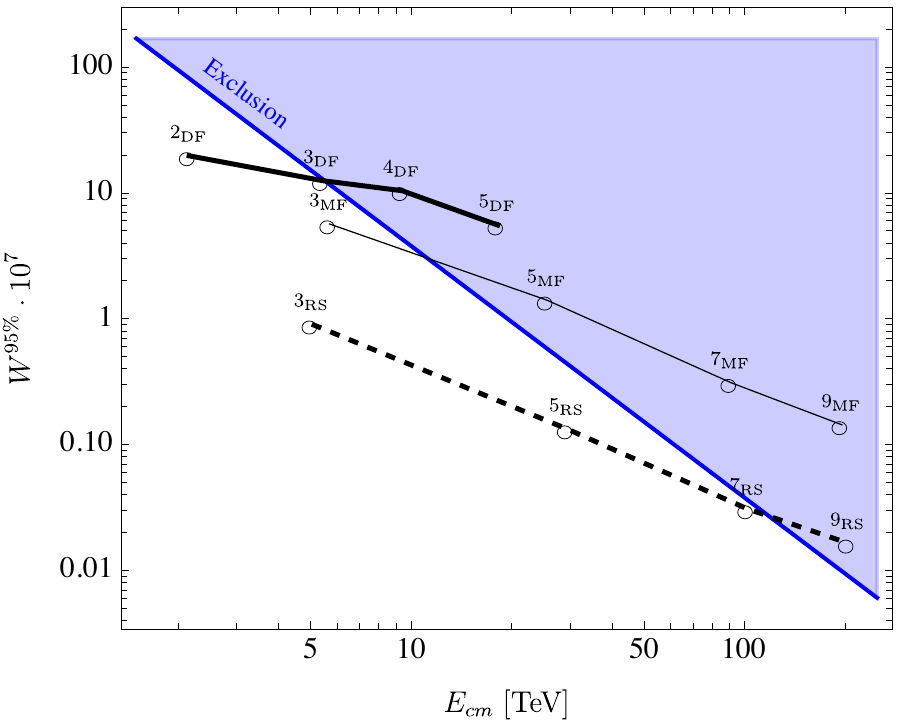}
\caption{The effect on the $W$ parameter from Dirac Fermions, Majorana Fermions,  and Real Scalars for different $n$ for thermal masses given in Tab.~\ref{tab:WIMP-thermal-mass}. The blue shaded area corresponds to the 95\% CL  exclusion on $W$ that can be attained at a muon collider running at center of mass energy $E_{cm}$.}
\label{fig:W}
\end{center}
\end{figure}

The fact that high energy muon colliders can probe quadratically smaller values of $W$ as the center-of-mass energy increases is a key to enable this sensitivity to heavy WIMPs. In fact, the value of $W$ generated by the WIMPs of Tab.~\ref{tab:WIMP-thermal-mass} becomes smaller as larger $n$ is considered, but it decreases less fast than the quadratic improvement of the bound on $W$. This is due to the fact that the effect on $W$ is suppressed by the mass of the WIMP on one hand and it is enhanced by $n$ on the other hand, i.e. $W\propto  C_{n,\textrm{eff}}/M^{2}$. For $M$ in Tab.~\ref{tab:WIMP-thermal-mass} roughly scales as $M\sim n^3$ and $C_{n,\textrm{eff}}\sim n^3$, we get  $W\sim 1/n^3\sim 1/M$ when the mass of the WIMP is fixed at the value predicted to be a thermal relic.  Thus we are lead to find that for a sufficiently large center-of-mass energy one or more WIMPs outside the kinematic reach of the collider can  be probed by the measurement of the $W$ parameter. The exact value of  $n$ at which WIMPs of different spin can be probed through $W$ can be obtained from Fig.~\ref{fig:W}, that is obtained assuming a collider with $E_{cm}=2M$ is used to search a WIMP of mass $M$. This assumption, though it may invalidate the EFT expansion, turns out to be conservative, as the   EFT computation of the effect of $\chi$ do not enjoy the threshold enhancement that is clearly observed in Fig.~\ref{fig:bounds-largen-scalars}.

\section{Conclusions}

The puzzle on the nature of the Dark Matter of the Universe is a very sound motivation to extend the Standard Model of particle physics. The search activity for the several proposed Dark Matter candidates is wide in scope and has been already a main subject of research for decades. A most motivated proposal for Dark Matter is that of a new matter field charged under $SU(2)$ weak interactions and, with suitable arrangements, possibly charged under $U(1)_{Y}$. 

A ``catalog'' of the possible  candidates of this type can be made by listing all the $n$-plets of weak $SU(2)$ for which the gauge charge carried by the $n$-plet does not spoil the perturbativity of the SM at energies too close to the mass of the $n$-plet itself. That is to say that the new theory made by adding   $\chi$ to the SM ought to be sufficiently perturbative to allow reliable thermal relic calculations and to imagine that such theory could be at least a valid effective field theory for a decade of energies. Concrete results about possible  Landau poles and transition scattering amplitudes find that candidates with $n\simeq 10$ are at best border-line, while candidates up to $n=7$ can be safely considered as WIMP candidates. Thermal masses for these candidates tend to saturate the order of magnitude estimate for the perturbative ``WIMP miracle'' usually quoted in the ballpark of 100~TeV. 

The mass scale of saturation of the perturbative limit for the ``WIMP miracle'' being so large has so far remained unreachable to direct and indirect collider probes. Even imagining a very large hadron collider such as a 100~km $pp$  100~TeV machine it would be just possible to scratch the surface in the search for WIMP candidates at the scale of saturation of perturbative unitarity. This is due to the fact that protons constituents can only reach a fraction of the $pp$ center-of-mass energy and, due to background processes, a hadron machine can reach WIMP candidates only up to $\mathcal{O}(5\%)$ of its beams center of mass energy~\cite{Cirelli:2014ai,Low:2014sh}. Thus it seems very hard, if not even impossible, to fully test the idea of WIMPs in collider experiments. 

Other search approaches for WIMPs in the lab comprise searches in ultra-clean underground experiments. These experiments can be sensitive to heavy WIMPs, although these are not in the best sensitivity mass range for the experiment. Very intriguingly, next generation experiments can potentially give signals pointing towards this direction. In principle also searches for new physics in high energy cosmic rays can give hints of WIMPs and future experiments can be sensitive to WIMPs up to masses close to the saturation of the perturbative limit discussed above. Searches in cosmic rays, however, are subject to significant uncertainties in the Dark Matter density profile in the object that hosts the Dark Matter. 

The perspectives for finding hints or evidence of WIMPs in   future experiments are encouraging, but the chance to produce WIMPs directly in the colliders and to study them in detail seems to require WIMPs to be quite light compared to the possible mass range they can span.

The situation at colliders is changed dramatically by the possibility to build high energy muon collider. Indeed a machine colliding point-like particle can exploit fully the beam energy to produce heavy states, such as heavy WIMPs. In addition, being a leptonic machine, the high energy muon collider promises to have a relatively clean collision environment, thus enabling precision measurements. 

We have explored the possibility to use precise measurements of fiducial cross-sections or  differential cross-sections to probe the existence of heavy WIMPs. We found that the amount of information that can be gained in differential studies is generally limited by the fact that the SM and new physics scattering amplitudes are very similar in the phase-space of the most abundant $2\to 2$ scatterings, thus motivating us to study mainly fiducial cross-section measurements. Such measurements can be carried out on a larger number of final states, as they do not require to tag electric charges of the final states, hence they can increase the mass reach of searches limited by the size  of the data sample. Indeed we find that including copious scatterings in all flavors of jets can improve the results appeared previously in the literature, which focused on final states for which the electric charges can be tagged, e.g. for the $b\bar{b}$ final state.  

In our study we have also included for the first time the effect of Dark Matter candidates in the production of neutral diboson final states $ZH$ and $WW$, and the effect of charged current scatterings in 3-body final states $Wf\bar{f}'$. These $Wf\bar{f}'$ final states are enhanced by the large energy of the collider that makes the emission of soft/collinear $W$ boson a very likely possibility, so much so that the computation of this scattering rate needs resummation. 

We find that running a muon collider at center of mass energy $E_{cm}$ and collecting integrated luminosity $10\,\mathrm{ab}^{-1}\cdot(E_{cm}/10 {\rm TeV})^{2}$ the study of the final states we included in our work is sufficient to exclude at 95\% CL all fermionic WIMP candidates in the WIMP catalog. For $E_{cm}\leq 14{\rm TeV}$ this amounts to a sensitivity to all fermionic WIMP candidates up to $n=5$, that is a Dark Matter candidate  with mass about 15~TeV. 

For scalars the expected signals are smaller and in general they require one order of magnitude more luminosity compared to same $n$ fermions in order to gain sensitivity through precision SM measurements. For scalar WIMPs light enough to be produced at colliders it is thus preferable to pursue a direct search strategy. Ref.~\cite{Bottaro:2022aa} finds that disappearing tracks and mono-X searches may lead to exclusions of all scalar WIMPs up to the $n=5$ complex scalar WIMP of around 11~TeV if muon colliders up to 30~TeV are considered, possibly with polarized beams or a luminosity within a factor few larger than the considered baseline. As per our Fig.~\ref{fig:bounds-largen-scalars}, exploiting polarization or a modest luminosity increase, a real scalar $n=5$, with a mass around 15~TeV, can leave observable deviations in precision observables. Therefore, putting these results together, we find that all scalar WIMPs up to $n=5$ can be probed at the muon collider running at energies up to 30~TeV. 

The energy up to 14~(30)~TeV  for muon colliders necessary to probe the entire catalog of fermionic (scalar) WIMPs up to $n=5$ can be considered ambitious. Nevertheless it is not an inconceivably large energy. This finding  brings heavy WIMPs in the realm of possible collider studies of dark matter and motivates research and development \cite{Black:2022ab,2203.07224v1,2203.08033v1,2203.07964v1,2203.07261v1,mucol2021,MICE2020,Delahaye:2019aa,Palmer_2014} for the realization of a high energy muon collider.

Very interestingly for the practical unfolding of   a possible collider program, the catalog of WIMPs offers a series of dark matter candidates that can be targets for lower energy stages of a multi-stages collider project. 

In addition, we have proven that for $2M>E_{cm}$ the EFT description matches well with our loop calculation. Thus, exploiting resummed limits~\cite{2202.10509v1} on the $W$ parameter for universal new physics, it is possible to compute expected limits at the muon collider on WIMP Dark Matter candidates up to the perturbative limit of the WIMP mass range around 100~TeV.

\section*{Acknowledgements}

It is a pleasure of RF to thank Salvatore Bottaro, Dario Buttazzo,
Marco Costa, Fabio Maltoni, Paolo Panci, Diego Redigolo, Lorenzo Ricci,  Ludovico Vittorio, and Andrea Wulzer for
  discussions and for collaborations on the physics of WIMPs and of the muon collider.
X. Zhao is supported by the Italian Ministry of Research (MUR) under grant PRIN 20172LNEEZ.

\appendix

\section{Calculation of the one-loop corrections}\label{app:loopfunc}
For the two-point self energy corrections,
the $I_A(s,M^2)$ for a Dirac Fermion is given by
\begin{align}
\begin{split}
    I_A^{F}(Q^2)=\frac{4}{3}[&-(1+2\frac{M^2}{Q^2})B_0(Q^2,M^2,M^2)\\
    &+2\frac{M^2}{Q^2}B_0(0,M^2,M^2)+\frac{1}{3}]\,,
    \end{split}
\end{align}
and for complex scalar it is given by
\begin{align}
\begin{split}
    I_A^{S}(Q^2)=\frac{1}{3}[&(4\frac{M^2}{Q^2}-1)B_0(Q^2,M^2,M^2)\\
    &-4\frac{M^2}{Q^2}B_0(0,M^2,M^2)-\frac{2}{3}]\,.
    \end{split}
\end{align}
The $B_0$ functions are in the standard convention \cite{Denner:1991kt},
and the explicit expression is given by
\begin{align}
\begin{split}
    B_0(Q^2,M^2,M^2)=&\frac{1}{\bar{\epsilon}}+2-\ln\frac{M^2}{\mu^2}\\
    &+\frac{\sqrt{x(x-4)}}{x}\ln\frac{-x+\sqrt{x(x-4)}}{x+\sqrt{x(x-4)}}\,,
    \end{split}
\end{align}
where $x=Q^2/(M^2-i0)$, $\mu$ is the renormalisation scale, and $\bar{\epsilon}^{-1}=2(4-d)^{-1}-\gamma_E+\ln 4\pi$ is the divergence in the $\overline{\textrm{MS}}$ convention.
In the low-energy limit $Q^2\ll M^2$, we have
\begin{align}
    B_0(Q^2,M^2,M^2)=\frac{1}{\bar\epsilon}-\ln\frac{M^2}{\mu^2}+\frac{x}{6}+\frac{x^2}{60}+\mathcal{O}(x^3),
\end{align}
and hence
\begin{align}
    I_A^F(Q^2)=&-\frac{4}{3}(\frac{1}{\bar\epsilon}-\ln\frac{M^2}{\mu^2})-\frac{4}{15}x+\mathcal{O}(x^2)\,,\\
    I_A^{S}(Q^2)=&-\frac{1}{3}(\frac{1}{\bar\epsilon}-\ln\frac{M^2}{\mu^2})-\frac{1}{30}x+\mathcal{O}(x^2)\,.
\end{align}
Clearly, the DM contribution to the gauge boson self-energies are UV-divergent,
and renormalisation is required.
Different from previous studies, here we adopt the on-shell renormalisation scheme, which properly takes into account the DM contribution to the EW input parameters $m_W,m_Z,G_{\mu}$.

Since \mgfive\, cannot handle modifications of the two-point Green functions natively, we absorb it into three point vertices.
In particular to deal with $f\bar{f}$ final states we add the following pieces to the SM $f\bar{f}V$ vertices for every SM fermion:
\begin{align}
    V^{\mu}[\bar{f}fZ]=&\gamma^{\mu}\bigg[(T^f_3\frac{g_2}{c_W}P_L-Q_f g_1s_W)\frac{1}{2}T_{ZZ}(p^2)\nonumber\\
    &\quad\ \,+Q_f g_2c_WT_{Z\gamma}(p^2)\bigg]\,,\\
    V^{\mu}[\bar{f}f\gamma]=&\gamma^{\mu}\left[Q_f g_2s_W\frac{1}{2}T_{\gamma\gamma}(p^2)\right]\,,
\end{align}
where $p^2=p_V^2$.
The renormalised self-energies are separated into weak part and hypercharge part, i.e. $T_{VV}(p^2)=T_{VV}^{W}(p^2)+T_{VV}^{Y}(p^2)$, and they are given by:
\begin{align}
    T_{ZZ}^W(p^2)=&\frac{g_2^2}{(4\pi)^2}F_W c_W^2\frac{p^2}{p^2-m_Z^2}[I_A(p^2)-I_A(m_Z^2)]\,,\\
    T_{Z\gamma}^W(p^2)=&\frac{g_2^2}{(4\pi)^2}F_W[s_W^2I_A(p^2)+c_W^2I_A(m_Z^2)-I_A(m_W^2)]\,,\\
    T_{\gamma\gamma}^W(p^2)=&\frac{g_2^2}{(4\pi)^2}F_W[s_W^2I_A(p^2)-\frac{c_W^4}{s_W^2}I_A(m_Z^2)\nonumber\\
    &\quad\quad\quad\quad \,+\frac{c_W^2-s_W^2}{s_W^2}I_A(m_W^2)]\,,\\
    T_{ZZ}^Y(p^2)=&\frac{g_2^2}{(4\pi)^2}F_Y \frac{s_W^4}{c_W^2}\frac{p^2}{p^2-m_Z^2}[I_A(p^2)-I_A(m_Z^2)]\,,\\
    T_{Z\gamma}^Y(p^2)=&\frac{g_2^2}{(4\pi)^2}F_Y\frac{s_W^4}{c_W^2}[-I_A(p^2)+I_A(m_Z^2)]\,,\\
    T_{\gamma\gamma}^Y(p^2)=&\frac{g_2^2}{(4\pi)^2}F_Ys_W^2[I_A(p^2)-I_A(m_Z^2)]\,,
\end{align}
where $c_W\equiv\cos\theta_W=\frac{m_W}{m_Z},s_W\equiv\sin\theta_W,F_W=\frac{1}{12}n(n^2-1),F_Y=nY^2$.
With the above effective vertices, the DM effects of all $\mu^+\mu^-\to f\bar{f}$ are included\footnote{Except the $\mu^+\mu^-\to\nu_{\mu}\bar{\nu}_{\mu}$ process which also receives contribution from the $W$ self-energy corrections, though it can be treated in an analogous manner.}.

To extend our analysis to diboson final states one should add further modifications of SM vertexes, e.g. $ZZH$, $ZWW$ and $\gamma WW$. We find instead more economical for diboson final states to introduce another implementation of the effects of $\chi$.  This is based on modifying only the vertex between muons and $Z/\gamma$:
\begin{align}
    V^{\mu}[\mu^-\mu^+Z]=&\gamma^{\mu}\bigg[(T^{\mu}_3\frac{g_2}{c_W}P_L-Q g_1s_W)T_{ZZ}(p^2)\nonumber\\
    &\quad\ \,+Q_{\mu} g_2c_WT_{Z\gamma}(p^2)\bigg]\,, \\
    V^{\mu}[\mu^-\mu^+\gamma]=&\gamma^{\mu}\bigg[(T^{\mu}_{3}\frac{g_2}{c_W}P_L-Q g_1s_W)\frac{p^2}{p^2-m_Z^2}T_{Z\gamma}(p^2)\nonumber\\
    &\quad\ \,+Q_{\mu} g_2s_WT_{\gamma\gamma}(p^2)\bigg] \,.
\end{align}
These muon vertices modifications can be used to account for all the (renormalised) self-energy corrections for $\mu^+\mu^-\to f\bar{f}\,(f\ne\mu)$ as well as $\mu^+\mu^-\to ZH$, and the $s$-channel self-energy corrections of $\mu^+\mu^-\to W^+W^-$. We stress that the process $\mu^+\mu^-\to W^+\mu^-\bar\nu_{\mu}$ also receives $t$-channel and $ZW$ fusion contributions that cannot be accounted within our method. This is why we have not included this final state in our analysis in Sec.~\ref{sec:mass-reach}.

To complete the calculation of $W^+W^-$ and $Wff'$ processes we also need to add the triangle diagram contributions, the $W$ wave-function renormalisation and the $W$ bubble correction from $\chi$. The $W$ wave-function and bubble correction to the propagator can be written in an analogous manner to the previously discussed bubble corrections to $Z$ and $\gamma$ propagators. For the triangle effects we compute diagrams using Qgraf\cite{Cullen:2011ac} and process them through FORM\cite{Vermaseren:2000nd}. The one-loop integrals are reduced to scalar integrals using Kira~\cite{Maierhofer:2017gsa}. The one-loop scalar integrals are evaluated by LoopTools~\cite{Hahn:1998yk} and the resulting lenghy expressions are added in the calculations as modfications of the tri-boson verteces in our \mgfive\, model.

\bibliographystyle{utphys}
\bibliography{bib.bib}   % name your BibTeX data base

%\begin{thebibliography}{}
% etc
%\end{thebibliography}

\end{document}